\documentclass[british,twocolumn,showpacs,preprintnumbers,amsmath,amssymb]{revtex4}
\usepackage[T1]{fontenc}
\usepackage[latin9]{inputenc}
\usepackage{amsmath}
\usepackage{graphicx}
\usepackage{amssymb}

\makeatletter

\usepackage{textcomp}

\@ifundefined{definecolor}
 {\usepackage{color}}{}

\makeatletter

\usepackage{dcolumn}

\usepackage{bm}

\makeatother

\makeatother

\usepackage{babel}

\begin{document}
\widetext

\title{s-wave superconductivity probed by measuring magnetic penetration
depth and lower critical field of MgCNi$_{3}$ single crystals}

\author{P. Diener$^{1}$, P. Rodière$^{1}$, T. Klein$^{1,2}$, C. Marcenat$^{3,4}$,
J. Kacmarcik$^{4}$, Z. Pribulova$^{4}$, D.J. Jang$^{5}$, H.S. Lee
$^{5}$, H.G. Lee$^{6}$, S.I. Lee$^{6}$}

\address{$^{1}$ Institut Néel, CNRS/UJF 25 rue des Martyrs BP 166 38042 Grenoble
cedex 9, FRANCE}

\address{$^{2}$ Institut Universitaire de France and Université Joseph Fourier,
B.P.53, 38041 Grenoble Cedex 9, France}

\address{$^{3}$ CEA, Institut Nanosciences et Cryogénie, SPSMS-LATEQS - 17
rue des Martyrs, 38054 Grenoble Cedex 9, France}

\address{$^{4}$ Safarik University, Slovak Academy Sciences, Institut of
Experimental Physics, Center for Low Temperature Physics, Kosice 04001,
Slovakia}

\address{$^{5}$ Department of Physics, Pohang University of Science and Technology,
Pohang 790-784, Republic of Korea}

\address{$^{6}$ National Creative Research Initiative Center for Superconductivity,
Department of Physics, Sogang University, Seoul, Korea}

\begin{abstract}
The magnetic penetration depth $\lambda$ has been measured in MgCNi$_{3}$
single crystals using both a high precision Tunnel Diode Oscillator
technique (TDO) and Hall probe magnetization (HPM). In striking contrast
to previous measurements in powders, $\delta\lambda$(T) deduced from
TDO measurements increases exponentially at low temperature, clearly
showing that the superconducting gap is fully open over the whole
Fermi surface. An absolute value at zero temperature $\lambda(0)=$230$\,$nm
is found from the lower critical field measured by HPM. We also discuss
the observed difference of the superfluid density deduced from both
techniques. A possible explanation could be due to a systematic decrease
of the critical temperature at the sample surface. 
\end{abstract}


\maketitle
The interplay between magnetism and superconductivity is currently
a subject of great interest. In the UGe$_{2}$ and URhGe uranium compounds,
for instance, a long range ferromagnetic ordered phase coexists with
the superconducting phase and a mechanism of spin fluctuations (SF)
could be at the origin of the Cooper pair formation \citep{Aoki2001,Saxena00}.
The recent discovery of high temperature superconductivity in oxypnictides
also rapidly became the topic of a tremendous number of both experimental
and theoretical works. The parent undoped LnOFeAs (where Ln=La,Sm,...)
compound is here close to itinerant magnetism due to the presence
of a high density of Fe d states at the Fermi level \citep{Singh_PRL08},
leading to competing ferromagnetic and antiferromagnetic fluctuations.
Similarly in the cubic (anti-)perovskite MgCNi$_{\text{3}}$ compound
\citep{He2001}, the presence of a strong Van Hove singularity in
the density of Ni states slightly below the Fermi level also leads
to strong ferromagnetic fluctuations \citep{DugdalePRB01,Rosner2001,SinghPRB01,Singer2001,Shan2005}.
These two systems have also a Fermi surface composed of both electron
and hole pockets (3D sheets in MgCNi$_{3}$ as compared to quasi-cylindrical
sheets in oxypnictides).

Despite these striking similarities in their electronic and magnetic
properties, spin fluctuations lead to very different effects in those
systems. On the one hand, ab-initio calculations rapidly showed that
the electron-phonon coupling constant ($\lambda_{e-ph}\sim0.2$) is
far too low to account for the high critical temperatures observed
in oxypnictides (up to $\sim55$ K) and an unconventional mechanism
mediated by the SF associated with a sign reversal of the (s-wave)
order parameter between electrons and holes sheets of the Fermi surface
has been proposed \citep{mazin_PRL08}. On the other hand, it has
been suggested that the narrow van Hove singularity could be responsible
for a nearly unstable phonon mode in MgCNi$_{3}$ inducing a large,
although reduced by SF, $\lambda_{e-ph}$ \citep{IgnatovPRB2003,Heid2004}
in agreement with experiments which yield an average electron-phonon
coupling constant in the order of 1.7 \citep{IgnatovPRB2003,Walte2004, Walte2005}.
The interplay between electron-phonon coupling and SF is further emphasized
in this system by the existence of a large isotopic effect \citep{Klimczuk2004}
which has been suggested to be enhanced by the strong SF \citep{Dolgov_PRL05}.

In this context, the nature of the superconducting order parameter
rapidly became a crucial issue. In MgCNi$_{3}$, the experimental
results still remain controversial : on the one hand, penetration
depth measurements (in polycrystalline samples) showed a quadratic,
i.e. non s-wave, temperature dependence suggesting a nodal order parameter
\citep{Prozorov2003}, whereas specific heat measurements clearly
indicate that the superconducting gap ($\Delta$) is fully open, with
a $\Delta/k_{B}T_{c}$ ratio ranging from $1.9$ to $2.1$ \citep{Lin2003,Shan2003,Walte2004}
i.e. well above the BCS weak coupling 1.76 value.

In this paper, we present high precision magnetic penetration depth
and lower critical field measurements performed in the same MgCNi$_{3}$
single crystals. We show that $\lambda(T)$ clearly follows an exponential
temperature dependence for $T<T_{c}/3$ showing that the gap is fully
open on the whole Fermi surface. A zero temperature $\lambda_{0}$
value of $230\,$nm, i.e. well above the London clean limit BCS value
($\sim60\,$nm) has been deduced from first penetration field measurements,
clearly suggesting the presence of strong mass renormalization and/or
impurity scattering effects. Introducing this value into the TDO data
however leads to a temperature dependence of the normalized superfluid
density $\frac{\rho_{S}(T)}{\rho_{S}(0)}=[\frac{1}{1+\delta\lambda(T)/\lambda_{0}}]^{2}$
which is different from the one directly deduced from the lower critical
field ($H_{c1}\propto ln(\kappa)/\lambda^{2}$ where $\kappa=\lambda/\xi$).
Possible reasons for this discrepancy are discussed.

\medskip{}

Single crystals were grown in a high pressure furnace as described
elsewhere \citep{LeeAdvMat07}. AC specific heat have been performed
on several samples of the same batch \citep{KacmarcikAPP08}. All
the measured crystals show sharp superconducting transitions ($\Delta T_{c}\sim0.2$$\,$K)
emphasizing the excellent bulk homogeneity of each crystal. We observed
however a large dispersion of critical temperatures from sample to
sample, between approximately 5.9 to 7.6$\,$K, probably due to a
slight Ni deficiency in the MgCNi$_{3}$ structure \citep{LeeAdvMat07}.
Three single crystals with a thickness of 0.1mm but different shapes
and critical temperatures have been selected. Sample \#A can be approximated
by a disk with a diameter of 0.3mm. Samples \#B and \#C have a rectangular
shape of 0.21$\times$0.15$\,$mm$^{2}$ and 0.24$\times$0.36$\,$mm$^{2}$
respectively. Samples \#A and \#B both present a bulk $T_{c}$ of
6.9$\,$K and exhibit exactly the same behavior by TDO and HPM, whereas
sample \#C has the highest $T_{c}$ at 7.6$\,$K.

The magnetic penetration depth has been measured with a high stability
LC oscillator operating at 14$\,$MHz, driven by a Tunnel Diode \citep{Degrift1975,Carrington1999}.
The AC excitation field is below 1$\,$\textmu{}T and the DC earth
magnetic field is screened by a demagnetized weak ferromagnet amorphous
ribbon, ensured to work well below $H_{c1}$. The sample stage, placed
at the bottom of a home-made He$^{3}$ refrigerator, is regulated
between 0.5$\,$K and 10$\,$K, whereas the LC oscillator remains
at fixed temperature. The superconducting sample is glued with vacuum
grease at the bottom of a sapphire cold finger, which can be extracted
in-situ \citep{Prozorov2000a}. The small filling factor of the excitation
coil by the superconducting sample ($\sim0.01\%$) ensures a small
perturbation of the circuit and the frequency shift $\delta f$ divided
by the one induced by the extraction of the superconducting sample,
$\Delta f_{\text{0}}$, is then proportional to the imaginary part
of the surface impedance and hence to the magnetic penetration depth\citep{Fletcher2007}.
As shown in the inset of Fig. 1, all samples present a sharp superconducting
transition at a critical temperature $T{}_{c}^{f}$ (defined by the
onset of the frequency shift change) equal to $6.9$$\,$K (resp.
$7.6$$\,$K) for sample \#A (resp. sample \#C) in good agreement
with the $C_{p}$ measurements.

\begin{figure}[h]
\begin{centering}
\includegraphics[clip,width=8cm]{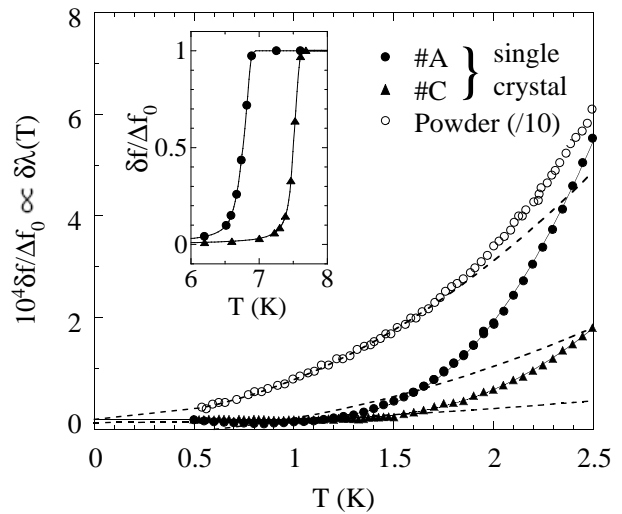}\label{fig1} 
\par\end{centering}

\caption{{\small Low temperature TDO frequency shift $\delta f$ normalized
by the frequency shift for a total extraction, obtained for single
crystals \#A and \#C. Open circles correspond to previous results
on polycrystalline powders \citep{Prozorov2003} divided by a factor
10. The dashed line is the $T^{2}$ law below 1.8$\,$K reported for
the powder. In the case of the single crystals, a better fit is obtained
with an exponential law (solid line). }\emph{\small Inset:}{\small {}
frequency shift at the critical temperature for both single crystals. }}

\end{figure}

Fig.1 displays the temperature dependence of the frequency shift,
proportional to $\delta\lambda(T)$, compared to the results previously
reported in powders \citep{Prozorov2003}. The amplitude of the shift
is 10 times larger in the case of the powder reflecting the fact that
the surface on which the supercurrents are flowing is much larger
in powders than in single crystals (for the same sample volume). It
is important to note that the temperature dependence of $\lambda$
is strikingly different in single crystals than in powder for which
a $T^{2}$ power law has been reported below 1.8$\,$K. Such a dependence
has been interpreted as an evidence for unconventional superconductivity
\citep{Prozorov2003} but our measurements do not support this scenario
as a $T^{2}$ power law only very poorly describes the experimental
data (see dashed line in Fig.1).

A very good fit to our data is actually obtained assuming the low
temperature approximation for clean type II superconductors with a
fully open gap : $\lambda(T)\propto$$\sqrt{\Delta/k_{B}T}$ e$^{-\Delta/k_{B}T}$.
This expression is valid for $k_{B}T<\Delta/5$, and leads to $\Delta/k_{B}=11.6(1)$$\,$K
for sample \#A (and B) and $\Delta/k_{B}=12.3(1)$$\,$K for sample
\#C. Note that this fitting procedure can lead to a slightly overestimated
$\Delta$ value (up to 10\%, depending on the range of the fit and
the ratio between $\Delta$ and $T_{c}$) but unambiguously shows
that the gap is fully open in good agreement with previous tunneling
spectroscopy \citep{Kinoda12001,Shan2003a}, NMR \citep{Singer2001},
and specific heat measurements which led to $\Delta/k_{B}\approx$13.0(2)$\,$K,
10.5$\,$K and 13$\,$K, respectively.

\medskip{}

However, fitting the low temperature data only leads to the size of
the minimum superconducting gap. To unambiguously exclude the presence
of any other gaps (and/or other gap symmetries) it is necessary to
analyze the full temperature dependence of the normalized superfluid
density $\rho_{S}(T)\propto1/\lambda(T)^{2}$ up to $T_{c}$. This
superfluid density can be deduced :

- either from the temperature dependence of the lower critical field
: $H_{c1}=\Phi_{0}^{2}/(4\pi\lambda^{2})(Ln(\kappa)+c(\kappa))$ where
$\kappa=\lambda/\xi$ ($\xi$ being the coherence length) and $c(\kappa)$
a $\kappa$ dependent function tending towards $\sim0.5$ for large
$\kappa$ values. As $\kappa$ is almost temperature independent (being
in the order of 40), $H_{c1}(T)$ is directly proportional to the
superfluid density which we will call $\rho_{S}^{Hc1}$.

- or by introducing the absolute value of the penetration depth at
T=0$\,$K ($\lambda_{0}$) into the TDO data : $\rho_{S}^{TDO}(T)\propto[\frac{1}{1+\delta\lambda(T)/\lambda_{0}}]^{2}=[\frac{1}{1+\delta f(T)/\Delta f_{0}\times R/\lambda_{0}}]^{2}$
where $R$ is a geometrical factor\citep{Prozorov2000b}.

\begin{figure}[h]

\begin{centering}
\includegraphics[clip,width=8cm]{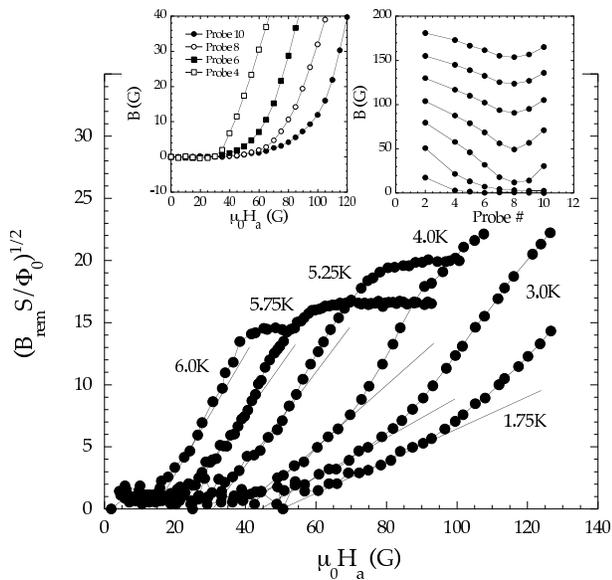} 
\par\end{centering}

\caption{{\small Remanent field $B_{rem}$ in flux quantum units ($\Phi_{0}/S$,
$S$ being the active area of the probe) as a function of the applied
field $H_{a}$ in sample \#A showing that $B_{rem}$ remains close
to zero up to $H_{a}=H_{p}$ (see text for details). }\emph{\small Left
inset:}{\small {} Local induction at $T=4.2$$\,$K as a function
of the applied field for several probe positions (see right inset)
showing that, even the penetration is much stronger close to the edges
(probe 4), the same first penetration field ($\sim$35$\,$G) can
be obtained on all of the probes. }\emph{\small Right inset:}{\small {}
Field profiles at $T=4.2\,$K for different values of the applied
field (measured on probe 2) clearly showing the Bean profile characteristic
of bulk pinning. Probe 8 is located close to the center of the sample
and probe 4 close to the sample edge. The spacing between probes is
$20\mu$m.}}

\end{figure}

The local magnetic induction has been measured with a miniature $16\times16$$\,$\textmu{}m$^{2}$
Hall probe and the first penetration field $H_{p}$ has been deduced
by measuring the remanent field ($B_{rem}$) in the sample after applying
an external field $H_{a}$ and sweeping the field back to zero. For
$H_{a}<H_{p}$ (i.e. in the Meissner state) no vortices penetrate
the sample and the remanent field remains equal to zero (actually
close to zero due to partial penetration through the sample corners).
$H_{a}$ is then progressively increased until a finite remanent field
is obtained (see Fig.2). Indeed, since vortices remain pinned in the
sample, $B_{rem}$ rapidly increases for $H_{a}>H_{p}$, varying as
$(H_{a}-H_{p})^{\alpha}$ with $\alpha=0.4\pm0.1$ (solid lines in
Fig.2 for $\alpha=0.5$). We get $H_{p}\sim50\pm5$$\,$G, $\sim55\pm5$$\,$G
and $\sim70\pm10\,$G for sample \#A, \#B and \#C, respectively. In
samples with rectangular cross sections, $H_{p}$ is then related
to $H_{c1}$ through $H_{c1}\approx H_{p}/tanh(\sqrt{\alpha d/2w})$
where $\alpha$ varies from 0.36 in strips to 0.67 in disks \citep{Brandt99,Zoldov94}.
Taking an average $\alpha$ value $\sim0.5$ we hence get $H_{c1}^{c}(0)\sim125\pm15\,$G
and correspondingly $\lambda_{0}=230\pm15$$\,$nm (introducing $H_{c2}(0)=\Phi_{0}/2\pi\xi(0)^{2}=9.5$$\,$T
\citep{KacmarcikAPP08}).

This value is in good agreement with the value deduced from muon spin
relaxation data in ceramics \citep{McDougall}, previous lower critical
field measurements in powder \citep{Lu2005} as well as the value
calculated from the thermodynamic critical field deduced from specific
heat measurements by Wälte \textit{et al.}

Note that, as pointed out by Wälte \textit{et al.}, this value is
much larger than the London clean limit BCS value $\lambda_{L}(0)=c/\omega_{p}$
$\sim60$$\,$nm ($\omega_{p}$ being the plasma frequency $\sim3.2$$\,$eV
\citep{Walte2004}). In the presence of strong mass renormalization
and/or impurity scattering effects $\lambda_{0}=\lambda_{L}(0)\sqrt{1+\lambda_{e-ph}}\sqrt{1+\xi_{0}/l}$
where $l$ is the mean free path and $\xi_{0}\sim\frac{\hbar v_{F}}{\pi\Delta_{0}}$
($v_{F}$ being the bare Fermi velocity). Introducing $\lambda_{e-ph}\sim1.8$
and $v_{F}\sim2.1\times10^{5}$$\,$m.s$^{-1}$ \citep{Walte2004},
one obtains $l\sim\xi_{0}/4\sim10$$\,$nm hence confirm that both
strong coupling and strong impurity diffusion are present. Note that
this $l$ value corresponds to a resistivity $\rho=v_{F}/\epsilon_{0}l\omega_{p}^{2}\sim10\,$\textmu{}$\Omega$cm
i.e. slightly lower than the residual resistivity measured in similar
crystals $\rho\sim30$$\,$\textmu{}$\Omega$cm \citep{LeeAdvMat07}.
However, since $\rho$ is expected to be in the order of $\lambda_{0}^{2}\mu_{0}\pi\Delta/\hbar(1+\lambda_{e-ph})$
(dirty limit), a residual resistivity of $30$$\,$\textmu{}$\Omega$cm
value would thus require that $\lambda_{e-ph}\ll1$ in striking contrast
with reported values.

\medskip{}

$\rho_{S}^{Hc1}$ (solid symbols) and $\rho_{S}^{TDO}$ (open symbols)
are displayed in Fig. 3 for sample \#A (squares) and \#C (circles).
The two techniques lead to strikingly different temperature dependence
for the superfluid density. For an isotropic superconducting gap,
the BCS superfluid density $\rho_{S}(T)$ reduced by thermally activated
excitations is expected to be given by : \begin{equation}
\rho_{S}(T)=1-\int\frac{\partial f}{\partial E}\frac{E}{\sqrt{E^{2}-\Delta^{2}(T)}}\label{eq1}\end{equation}
 where $f$ is the Fermi Dirac distribution, $E$ the energy above
the Fermi energy, $\Delta(T)$ the value of the superconducting gap
at the temperature $\mathit{T}$. As shown in Fig.3 (solid lines)
very good fits to the $\rho_{S}^{Hc1}$ data are obtained using an
alpha model in which the temperature dependence of the superconducting
gap (normalized to its $\mathit{T}$=0$\,$K value) has been assumed
to be equal to the reduced BCS weak coupling value calculated from
the gap equation \citep{Padamsee1973} and taking $\Delta(0)=2k_{B}T_{c}$.
Note that a superconducting gap equal to its weak coupling theory
value ($\Delta(0)=1.76k_{B}T_{c}$) only leads to a poor fit of the
data, confirming the large value of the $\Delta(0)/k_{B}T_{c}$ ratio
previously obtained by bulk probes such as specific heat measurements.

\begin{figure}
\begin{centering}
\includegraphics[clip,width=9cm]{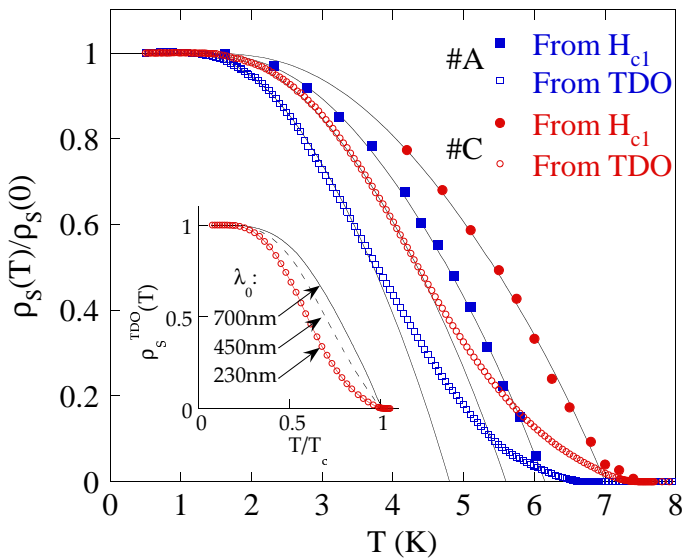} 
\par\end{centering}

\caption{{\small (Color online) Normalized superfluid density deduced from
H$_{c1}$ measurements (full symbols) and TDO measurements with $\lambda_{0}=$230$\,$nm
(open symbols) for samples \#A (blue squares) and \#C (red circles).
The solid lines are the fit for a superconducting gap $\Delta=2\, k_{B}T_{C}$
with $\mathit{T_{c}=}$4.8, 5.6, 6.2 and 7$\,$K (see text). }\emph{\small Inset:}{\small {}
influence of the $\lambda_{0}$ value used to deduce $\rho_{S}$ from
TDO measurements. }}

\end{figure}

On the other hand, $\rho_{S}^{TDO}$ displays a strong downward curvature
at low temperature followed by a clear upward curvature as the superfluid
density drops below 0.5 (i.e. for $\lambda(T)>1.4\lambda_{0}$). As
pointed out above one has to determine the $R/\lambda_{0}$ ratio
in order to deduce $\rho_{S}^{TDO}$ from the $\delta f/\Delta f_{0}$
data. The $R$ value has been calculated from the aspect ratio using
the formula introduced by Prozorov \citep{Prozorov2000b}. The validity
of this procedure has been checked on Pb samples. Moreover, different
AC magnetic field orientations on the same single crystal of MgCNi$_{3}$(but
different R) show the same quantitative temperature dependence of
$\lambda(T)$, consistently with an isotropic cubic system.

A possible explanation would hence be an underestimation of $\lambda_{0}$.
The influence of $\lambda_{0}$ is displayed in the inset of Fig.3.
As shown taking $\lambda_{0}\simeq700\,$nm instead of 230$\,$nm
leads to a temperature dependence for $\rho_{S}^{TDO}$ similar to
the one obtained for $\rho_{S}^{Hc1}$. This value is however well
above our error bars on $\lambda_{0}$ and would correspond to $\mu_{0}H_{c1}(0)\sim15\,$G
i.e. even smaller than our first penetration field values ($\sim55\,$G).
Note that strong bulk pinning could lead to an overestimation of $H_{p}$,
if measured in the center of the sample (see for instance \citep{okazaki_PRB09})
but we checked that very similar $H_{p}$ values are obtained for
several probe positions by placing the sample on an array of 11 miniature
($10\times10\,$\textmu{}m$^{2}$) probes : as shown in the left
inset of Fig.2, the field distribution clearly presents the $V-shape$
profile characteristic of a strong bulk pinning. Even though those
profiles confirm the good homogeneity of the sample, one can not exclude
the presence of a strong disorder at the surface of the samples leading
to a \textit{surface} penetration field much larger than the bulk
value. However, the $\lambda_{0}$=$700\,$nm value would require
an extremely small mean free path ($\sim1\, nm$, see discussion above).
A possible difference between the mixed state and Meissner state penetration
depth values associated either to a Doppler shift induced by the supercurrents
on the excitation spectra \citep{Yip1992,YipXu1995} or to a strong
field dependence of $\lambda$ in the mixed state (see for instance
\citep{klein_PRB06}) due to multiband effects can be excluded in
our isotropic, fully gapped system.

Another explanation could be a difference between $\mathit{bulk}$
and $\mathit{surface}$ critical temperature. Indeed, at low temperature
TDO measurements only probe the sample on a typical depth in the order
of $\lambda_{0}\sim0.2\,$\textmu{}m, i.e. roughly 0.4\% of the
total volume (for a Volume to Surface ratio of 50$\,$\textmu{}m).
In the presence of a weak coupling superconducting gap, this volume
only increases to about 20\% of the sample volume for $T\rightarrow T_{c}/2$
and the bulk of the sample is only probed close to $T_{c}$ as the
magnetic penetration depth finally diverges for $T\rightarrow T_{c}$.
On the other hand, the Hall probe has been placed close to the center
of the sample in the HPM measurements and is hence sensitive to the
$\mathit{bulk}$ of the sample. In the case of MgCNi$_{3}$, it is
known that the critical temperature has a surprising high sensitivity
to a very small change in the C or Ni stoechiometry \citep{He2001, LeeAdvMat07}
and also surface stress \citep{Uehara2006,Walte2005}. Assuming that
the critical temperature of the surface is 20$\,$\% smaller than
the bulk value, very good fit to the data could be obtained for $\rho_{S}^{TDO}$
using Eq.1 for $\mathit{T<\frac{3}{4}T_{c}}$ (still taking $\Delta(0)=2k_{B}T_{c}$,
see solid lines in Fig.3). Note that a large dispersion of the $T_{c}$
values in powder might explain the anomalous temperature dependence
observed in previous $\lambda$ measurements. Clear deviations from
the standard BCS theory (Eq.1) have been observed in systems like
MgB$_{2}$ \citep{Golubov2002} or more recently in pnictides \citep{gordon_PRL09}
but in our case those deviations in $\rho_{S}^{TDO}$ are due to surface
inhomogeneities (disorder and/or $T_{c}$) and our measurements emphasize
the importance of coupling complementary experimental probes in order
to unambiguously address this issue.

\medskip{}

To conclude, we have shown that the temperature dependence of the
magnetic penetration depth is exponential in MgCNi$_{3}$ single crystals
signalling the presence of a fully open superconducting gap. A drastically
different behavior has systematically been observed between the superfluid
density extracted from the lower critical field and TDO measurements
performed on the same sample, which are most probably due to surface
disorder and/or a depletion of 20\% of the critical temperature at
the surface.

We are most obliged to V. Mosser of ITRON, Montrouge, for the development
of the Hall sensors used in this study. This work was partially supported
by the Slovak R\&D Agency under Contracts No. VVCE-0058-07, No. APVV-0346-07,
and No. LPP-0101-06.

\bibliographystyle{apsrev}

\end{document}